\documentclass[pra,twocolumn,showpacs,tightenlines]{revtex4-1}
\usepackage{graphicx}
\usepackage{braket}
\usepackage{amsmath}
\usepackage{bm}
\begin{document}

\title{Non-Markovian finite-temperature 
two-time correlation functions of system operators of
%beyond the quantum regression theorem:
a pure-dephasing model}
\author{ Hsi-Sheng Goan}\email{goan@phys.ntu.edu.tw}
\author{Chung-Chin Jian}
\author{Po-Wen Chen}
\affiliation{Department of Physics  and Center for Theoretical
  Sciences, National Taiwan University, Taipei 10617, Taiwan and\\
Center for Quantum Science and Engineering, National Taiwan University, Taipei 10617, Taiwan}
\date{\today }

\begin{abstract}
We evaluate the non-Markovian finite-temperature 
two-time correlation functions (CF's) of system operators of a pure-dephasing spin-boson model
in two different ways, one by the direct exact operator technique and the other by the recently derived evolution equations, valid to second order in the system-environment interaction Hamiltonian. 
This pure-dephasing spin-boson model that is exactly solvable has been extensively studied as a simple decoherence model.
However, its exact non-Markovian finite-temperature 
two-time system operator CF's, to our knowledge, have not been presented in the literature. This may be mainly due to the fact, illustrated in this article, that in contrast to the Markovian case, the time evolution of the reduced density matrix of the system (or the reduced quantum master equation) alone is not 
sufficient to calculate the two-time system operator CF's of  
non-Markovian open systems.
The two-time CF's obtained using the recently derived evolution equations 
in the weak system-environment coupling case for this non-Markovian 
pure-dephasing model 
happen to be the same as those obtained from the exact evaluation.
However, these results significantly differ from the non-Markovian two-time CF's obtained by wrongly directly applying the quantum regression theorem (QRT), a useful procedure to calculate the two-time 
CF's for weak-coupling Markovian open systems. 
This demonstrates clearly that the recently derived evolution equations 
generalize correctly the QRT to non-Markovian finite-temperature cases.
It is believed that these evolution equations 
will have applications in many different branches of physics.
\end{abstract}

\maketitle

\section{INTRODUCTION}
A quantum system is inevitably subject to the influence of its surroundings or environments \cite{Scully97, Carmichael99,Gardiner00,Milburn08,Paz01,Breuer02}. An environment usually consists of a practically infinite number of degrees of freedom and acts statistically as a whole identity referred as a reservoir or bath of the open quantum system. 
Most often, one is concerned with only the system dynamics and the key quantity is the reduced system density matrix $\rho(t)$ defined as the partial trace of the total system-plus-reservoir density operator $\rho_T(t)$ over the reservoir degrees of freedom; i.e., $\rho(t)={\rm Tr}_R[\rho_T(t)]$.
If the time evolution of the reduced density matrix that can be Markovian or non-Markovian is known, one is able to calculate the (one-time) expectation values or quantum average of the physical quantities of the system operators. 
But knowing the time evolution of the reduced density matrix is not sufficient 
to calculate the two-time (multiple-time) correlation functions (CF's) 
of the system operators in the 
non-Markovian case \cite{Alonso05,Vega06,Alonso07}. 

In the Markovian case, an extremely useful procedure to calculate the two-time (multiple-time) CF's is the so-called quantum regression theorem (QRT) \cite{Scully97,Carmichael99,Gardiner00,Milburn08} that gives
 a direct relation between the time evolution equation of the single-time expectation values and that of their corresponding two-time (multiple-time) CF's. 
So knowing the time evolution of the system reduced density matrix allows one to calculate all of the two-time (multiple-time) Markovian CF's. 
For the non-Markovian case, it is known that the QRT is not valid in general \cite{Ford96,Ford99,Lax00,Ford00}.  
Recently, using the stochastic Schr\"{o}dinger equation approach and the Heisenberg equation of system operator method, an evolution equation, valid to second order in system-environment coupling strength, 
%from which a set of coupled differential equations can be obtained 
for the two-time (multiple-time) CF's of the system operators has been derived for an environment at the zero temperature and for a system in an initial pure state \cite{Alonso05,Vega06,Alonso07}. 
This evolution equation has been applied to calculate the emission spectra of a two-level atom placed in a structured non-Markovian environment (electromagnetic fields in a photonic band-gap material) \cite{Vega08}.
In Ref.~\cite{Vega06},  an evolution equation for the reduced propagator of the system state vector, conditioned on an initial state of the environment differing from the vacuum, was derived using the stochastic Schrodinger equation approach. It is thus possible to use the reduced propagator to evaluate the expectation values and CF's of the system observables 
for general environmental initial conditions, 
not necessarily an initial vacuum state for the environment \cite{Vega06}.
%a master equation for calculating the time evolution of mean values 
%for general initial conditions, not necessarily an initial vacuum state 
%for the environment, has also been obtained using the stochastic 
%Schr\"{o}dinger equation approach.  
By using another
commonly used open quantum system technique, the quantum master
equation approach \cite{Scully97,
  Carmichael99,Gardiner00,Milburn08,Paz01,Breuer02},
we are able to extend the two-time CF evolution equation to a non-Markovian
finite-temperature environment for any initial system-environment
separable state.
%An extension of the two-time CF evolution equation to a non-Markovian
%finite-temperature environment for any initial system-environment
%separable state is  using another
%commonly used open quantum system technique, the quantum master
%equation approach \cite{Scully97,
%  Carmichael99,Gardiner00,Milburn08,Paz01,Breuer02}.
The detailed derivation will be presented elsewhere \cite{Goan09} but the 
essential results will be summarized in Sec.~\ref{two-time_EQ}. 
The derived evolution equation that generalizes the QRT to the non-Markovian finite-temperature case is believed to have applications in
many different branches of physics. 

The purpose of this article is twofold: (a) We show that in general the time evolution of the reduced density matrix of the system (or the reduced quantum master equation) alone is not sufficient to calculate the two-time CF's of the system operators of non-Markovian open systems, 
even in the weak system-environment coupling case. 
We present an evaluation of an exactly solvable non-Markovian model,
i.e., a pure-dephasing spin-boson model
\cite{Vega06,Unruh95,Palma96,Duan98,Diosi98,Reina02,Schaller08}, 
to justify the statement. The exact non-Markovian finite-temperature two-time CF's of the system operators of this model, to our knowledge, have not been presented in the literature.
(b) This exactly solvable model allows us to test the validity of the
derived non-Markovian finite-temperature evolution equation of
two-time CF's presented in Sec.~\ref{two-time_EQ}.
%in Ref.~\cite{Goan09}. 
%It is shown that the derived non-Markovian evolution equations, 
%with results of the two-time CF’s coinciding with those exact results 
%of this solvable model, correctly generalize the QRT to non-Markovian 
%finite-temperature cases.
It will be shown that the two-time CF's obtained using the 
evolution equation in the weak system-environment coupling 
limit \cite{Goan09}
in Sec.~\ref{two-time_EQ} 
for the exactly solvable non-Markovian model 
%in which ${H_S,L]=[H_S,L^{\dagger}]=0$ (pure-dephasing model), 
happen to be the same as those obtained from the exact evaluation.
%where $H_s$ denotes the system Hamiltonian and 
%$L$ stands for the system operator coupled to the environment.
However, these results significantly differ from the 
non-Markovian CF's obtained by wrongly applying directly the QRT.  
This demonstrates clearly 
that the derived evolution equations generalize correctly
the QRT to non-Markovian finite-temperature cases.

The article is organized as follows. We first
summarize the important results of the newly obtained 
evolution equations \cite{Goan09} that generalizes the QRT to 
the non-Markovian finite-temperature case  
in Sec.~\ref{two-time_EQ}.
After brief description of the pure-dephasing spin-boson model in the
beginning of Sec.~\ref{sec:Exact}, we calculate the exact time
evolution of the reduced density matrix of the system and one-time
expectation values 
in Sec.~\ref{sec:RDM}.   
The exact two-time CF's are evaluated in subsection \ref{sec:Exact_2TCF}.
In Sec.~\ref{derived_EE}, we use the derived evolution equations in  Ref.~\cite{Goan09} to calculate the one-time and two-time CF's. It is shown that the results obtained in Sec.~\ref{derived_EE} are the same as those by the exact evaluation in Sec.~\ref{sec:Exact}. This demonstrates the validity and practical usage of the derived evolution equations in Ref.~\cite{Goan09}.  Numerical results and discussions are presented in Sec.~\ref{sec:result}. A short conclusion is given in Sec.~\ref{sec:conclusion}.

\section{Evolution equation of non-Markovian finite-temperature
  two-time CF's}
\label{two-time_EQ}

A class of systems considered in \cite{Alonso05,Vega06,Alonso07}
is modeled by the Hamiltonian 
\begin{eqnarray}
H&=&H_{S}+H_I+H_R \nonumber \\
&=&H_S+\sum_{\lambda }\hbar g_{\lambda }\left( L^{\dagger} a_{\lambda}+La_{\lambda }^{\dagger}\right) 
+\sum_{\lambda }\hbar\omega_{\lambda}a_{\lambda }^{\dagger}a_{\lambda},
\label{BosonModel}
\end{eqnarray}
 where $H_{S}$ and $H_{R}$ are system and environment Hamiltonians, respectively, and $H_{I}$ stands for
the Hamiltonian that describes the interaction between the system and 
the environment.
%$H_S$ denotes the system Hamiltonian,
So 
%is the system operator coupled to the environment,
$L$ acts on the Hilbert space of the system, 
$a_{\lambda}^{\dagger}$ and $a_{\lambda}$ are creation and annihilation operators on the environment Hilbert
space, and $g_{\lambda}$ and $\omega_{\lambda}$ are the coupling strength and the frequency of the 
$\lambda$th environment oscillator, respectively. 
The derivations of the non-Markovian evolution equations of the 
two-time (multitime) CF's for the general Hamiltonian model (\ref{BosonModel})
in Refs.~\cite{Alonso05,Vega06,Alonso07} 
(Eq.~(6) in Ref.~\cite{Alonso05}, Eq.~(31) in Ref.~\cite{Vega06} and Eq.~(60) in Ref.~\cite{Alonso07}) are presented 
for an environment at the zero temperature and for a system state in
an initial pure state. 
It was mentioned in Ref.~\cite{Vega06} that it is possible to use the reduced stochastic system propagator that corresponds to an initial state of the environment different from the vacuum to evaluate the single-time expectation values and multitime CF's with more general initial conditions.  But only a master equation that is conditioned on initial bath states and is capable of evaluating the {\em single-time} expectation values of system observables for general initial conditions, both for an initial pure state and mixed state, 
was derived \cite{Vega06}. 
In Refs.~\cite{Alonso05,Vega06,Alonso07}, calculations of the {\em two-time} CF's of system observables for dissipative spin-boson models in thermal baths are, however, presented even though in their derivations of the {\em two-time (multitime)} evolution equations, the bath CF's are given in its zero-temperature form. This is possible due to the reason that for a system-environment model with a Hermitian system operator $L=L^\dagger$ coupled to the environment, the linear finite-temperature stochastic Schr\"odinger equation could be written in a simple form of the zero-temperature equation \cite{Diosi98,Yu04} if the zero-temperature bath CF is replaced with its corresponding effective finite-temperature bath CF. As a result, the evolution equation of thermal two-time (multitime) CF's for a Hermitian coupling operator $L=L^\dagger$ also becomes equal to its zero-temperature counterpart with the replacement of the zero-temperature bath CF with its effective finite-temperature bath correlation kernel. It is for this reason that the dissipative spin-boson model with a thermal environment can be studied with the two-time (multitime) evolution equations derived in 
Refs.~\cite{Alonso05,Vega06,Alonso07}, 
since in that model $L=\sigma_x=L^\dagger$. 
But this reduction of the finite-temperature evolution equation to its zero-temperature form \cite{Alonso05,Vega06,Alonso07} is not valid for more general non-Markovian finite-temperature cases where the system coupling operators are not Hermitian, i.e., $L\neq L^\dagger$. 
In other words,  if the system operator coupled to the environment is not Hermitian $L\neq L^\dagger$,
the two-time (multitime) differential evolution equations presented in 
Refs.~\cite{Alonso05,Vega06,Alonso07} are valid for a zero-temperature
environment only.

By using another
commonly used open quantum system technique, the quantum master
equation approach \cite{Scully97,
  Carmichael99,Gardiner00,Milburn08,Paz01,Breuer02},
it is possible to obtain in the weak system-environment coupling limit a  two-time evolution equation {\em for non-Markovian finite-temperature environments with both Hermitian and non-Hermitian system coupling operators and for any initial system-environment separable states}. 
The detailed derivation will be presented elsewhere \cite{Goan09} but the 
important results are summarized here.
%For the sake of completeness, we rewrite here 
%The relevant evolution equations derived in Ref.~\cite{Goan09}.   
The second-order evolution equations of the single-time expectation
values for the class of systems modeled by the Hamiltonian (\ref{BosonModel})
is
\begin{eqnarray}
&&
{d\left\langle A\left( t_{1}\right) \right\rangle }/{dt_{1}} \notag \\
&=&({i}/{\hbar }){\rm Tr}_{S}\left( \left\{[H_{S},A]\right\}(t_1) \rho(0) \right)\notag \\
&&+\int_{0}^{t_{1}}d\tau {\rm Tr}_S\notag \\
&&\quad \left( \alpha^{\ast }( t_{1}-\tau)\left\{\tilde{L}^{\dagger}(\tau -t_{1})[{A},{L}]\right\}( t_{1}){\rho}(0)\right.\notag \\
&&\quad+\alpha( t_{1}-\tau )\left\{ [ {L}^{\dagger},A] \tilde{L}(\tau -t_{1})\right\}(t_{1})\rho(0)\notag \\
&&\quad+\beta^\ast(t_{1}-\tau)\left\{\tilde{L}(\tau -t_{1})[A,L^{\dagger}]\right\}(t_{1})\rho(0)\notag \\
&&\quad+\left.\beta(t_{1}-\tau)\left\{[L,A] \tilde{L}^{\dagger}(\tau -t_{1})\right\}(t_{1})\rho(0)\right), 
\label{1time_evol_eq_f}
\end{eqnarray}
and that of the two-time CF's can be obtained as
\begin{eqnarray}
&&{d\left\langle A\left( t_{1}\right) B\left( t_{2}\right)\right\rangle }/{dt_{1}} \notag \\ 
&=&({i}/{\hbar}){\rm Tr}_{S}\left(\left\{[H_{S}, A]\right\}(t_{1}){B}(t_{2})\rho(0)\right) \notag\\
&&+\int_{0}^{t_{1}}d\tau {\rm Tr}_S\notag \\
&&\quad\left(\alpha^{\ast }( t_{1}-\tau)\left\{\tilde{L}^{\dagger}(\tau -t_{1})[{A},{L}]\right\}( t_{1}){B}(t_{2}){\rho}(0) \right.\notag \\
&&\quad+\alpha(t_{1}-\tau )\left\{[{L}^{\dagger},A]\tilde{L}(\tau -t_{1})\right\}(t_{1})B(t_{2})\rho( 0) \notag \\
&&\quad+\beta^{\ast }(t_{1}-\tau)\left\{\tilde{L}(\tau -t_{1})[A,L^{\dagger}]\right\}(t_{1}) B(t_{2}) \rho(0)\notag \\
&&\quad+\left.\beta(t_{1}-\tau)\left\{[L,A] \tilde{L}^{\dagger}(\tau -t_{1})\right\}(t_{1})B(t_{2})\rho(0) \right) \notag \\
&&+\int_{0}^{t_{2}}d\tau {\rm Tr}_S \notag \\
&&\quad\left( \alpha(t_{1}-\tau)\left\{[ L^{\dagger},A]\right\}(t_{1})\left\{[B,\tilde{L}(\tau -t_{2})]\right\}(t_{2})\rho(0) \right.\notag \\
&&\hspace{-0.8cm}+\left.\beta(t_{1}-\tau)\left\{[L,A]\right\}(t_{1})\left\{[B,\tilde{L}^{\dagger}(\tau-t_{2})]\right\}(t_{2})\rho(0) \right).
\label{2time_evol_eq_f}
\end{eqnarray}
Here
$\tilde{L}(t)=\exp\left(iH_St/\hbar\right)L\exp\left(-iH_St/\hbar\right)$
is the system operator in the interaction picture with respect to
$H_S$, and 
\begin{eqnarray}
%\alpha_{\rm eff}(\tau-s)&=&\alpha(\tau-s)+\beta(\tau-s), \label{alpha_eff}\\ 
\alpha(\tau-s)&=&\sum_{\lambda}(\bar{n}_{\lambda}+1)|g_{\lambda}|^2 
e^{-i\omega_{\lambda}(\tau-s)}, \label{alpha}\\
\beta(\tau-s)&=&\sum_{\lambda}\bar{n}_{\lambda}|g_{\lambda}|^2 
e^{i\omega_{\lambda}(\tau-s)}. \label{beta} 
\end{eqnarray}
%Note that $\alpha(\tau-s)$ and $\beta(\tau-s)$ 
are known as the environment CF's:
$\alpha(\tau-s)=\left\langle \sum_{\lambda}g_{\lambda}\tilde{a}_{\lambda}(\tau)\sum_{\lambda'}g_{\lambda'}\tilde{a}^{\dagger}_{\lambda'}(s)\right\rangle$ and $\beta(\tau-s)=\left\langle \sum_{\lambda}g_{\lambda}\tilde{a}^{\dagger}_{\lambda}(\tau)\sum_{\lambda'}g_{\lambda'}\tilde{a}_{\lambda'}(s)\right\rangle$, where $\tilde{a}_{\lambda}(\tau)=a_{\lambda}e^{-i\omega_\lambda \tau}$ and  $\tilde{a}^{\dagger}_{\lambda}(\tau)=a^{\dagger}_{\lambda}e^{i\omega_\lambda \tau}$ are the reservoir operators in the interaction picture.

We note here that 
for a Hermitian coupling operator $L=L^{\dagger}$ 
%as in the pure-dephasing model [see Eq.~(\ref{SpinBoson})]  
%presented here, 
the finite-temperature evolution equations (\ref{1time_evol_eq_f}) and (\ref{2time_evol_eq_f}) reduce,  respectively,    
to their zero-temperature counterparts 
but with the effective bath CF given by $\alpha(t_1-\tau)+\beta(t_1-\tau)$
\cite{Alonso05,Vega06,Alonso07}. 
This was pointed out to occur in general for $N$-time CF's  
in  Refs.~\cite{Alonso05,Vega06,Alonso07}.
%As a result, the pure-dephasing spin-boson model with $L=\sigma_z=L^{\dagger}$
%in a thermal environment can be studied using the zero-temperture evolution 
%equation \cite{Alonso05,Vega06,Alonso07,Goan09}. The evolution 
%equations (\ref{1time_evol_eq_f}) and (\ref{2time_evol_eq_f}) for 
%two-time thermal CF's derived in \cite{Goan09} are however applicable 
%to more general cases. 
%The correct reduction of the general evolution 
%equations (\ref{1time_evol_eq_f}) and (\ref{2time_evol_eq_f}) 
%to their zero-temperature forms with the effective bath CF given 
%by $\alpha(t_1-\tau)+\beta(t_1-\tau)$ for special Hermitian coupling 
%operator $L=L^{\dagger}$ models, nevertheless,
%constitutes a simple direct check of their validity. 

\section{Exact evaluations of pure dephasing spin-boson model}
\label{sec:Exact}

%non-Markovian evolution equations \cite{Goan09}
Here we consider an exactly solvable pure dephasing model of
\begin{equation}
H_S=(\hbar\omega_S/2)\sigma_z, \quad\quad
L=\sigma_z=L^{\dagger}  
\label{SpinBoson}  
\end{equation}
to test the evolution equations
(\ref{1time_evol_eq_f}) and (\ref{2time_evol_eq_f}).
This pure dephasing spin-boson model in which $[H_S,L]=0$ has been extensively studied as a simple decoherence model in the literature
\cite{Vega06,Unruh95,Palma96,Duan98,Diosi98,Reina02,Schaller08}. 
But most of the studies focus on the discussion of the time evolution of the reduced density matrix of the spin, or other one-time expectation values of the spin system operators. Recently, the two-time CF's of the system operators at the zero temperature for this model was reported in  Ref.~\cite{Vega06}. 
%but the results in \cite{Vega06} may contain some typos. 
%So to set the record straight and also 
Nevertheless, to demonstrate the validity and practical usage of the
finite-temperature non-Markovian evolution equation
of the two-time CF's (\ref{2time_evol_eq_f}), 
we present a detailed evaluation of the exact finite-temperature two-time CF's for this simple model. These exact non-Markovian finite-temperature two-time CF's of the system operators, to our knowledge, have not been presented in the literature. 

\subsection{Reduced density matrix and one-time expectation values}
\label{sec:RDM}
Before we derive the two-time CF's, we evaluate the exact time evolution of the reduced density matrix and one-time expectation values
%that have been reported in the literature 
for the non-Markovian spin-boson model.
%It is convenient to go to the interaction picture. 
In the interaction picture, the total density matrix of the combined (spin plus bath) system at time $t$ is given by 
\begin{equation}
\tilde{\rho}_{T}(t)=\tilde{U}(t)\rho_{T}(0)\tilde{U}^{\dagger}(t),
\label{eq:rhoTI}
\end{equation}
where the time evolution operator is 
\begin{eqnarray}
 \tilde{U}(t)&=&
%{\exp}\left(iH_{0}t/\hbar\right){\exp}\left(-iHt/\hbar\right)
e^{iH_{0}t/\hbar}e^{-iHt/\hbar}
\nonumber \\
&=&
{\rm T}\left[
e^{({-i}/{\hbar})\int_{0}^{t}d\tau \tilde{H}_{I}(\tau)}\right]. 
%\exp\left(\frac{-i}{\hbar}\int_{0}^{t}d\tau H_I(\tau)\right)\right]. 
\label{eq:evolution_op}
\end{eqnarray}
Here $H_{0}=H_{S}+H_{R}$, $\tilde{H}_{I}\left(t\right)={\rm exp}\left(iH_{0}t/\hbar\right)H_{I}{\rm exp}\left(-iH_{0}t/\hbar\right)$ and 
 ${\rm T}$ is the time-ordering operator
which arranges the operators with the earliest times to the right.   
>From Eqs.~(\ref{BosonModel}) and (\ref{SpinBoson}), 
a simple calculation gives
\begin{equation}
\tilde{H}_{I}(t)=\sum_{\lambda}\hbar g_{\lambda}\sigma_z
\left(e^{i\omega_{\lambda}t}a^{\dagger}_{\lambda}+e^{-i\omega_{\lambda}t}a_{\lambda}\right).
%\left( \exp\left(i\omega_{\lambda}t\right)a^{\dagger}_{\lambda}
%+\exp\left(-i\omega_{\lambda}t\right)a_{\lambda}\right).
\label{eq:HIt}
\end{equation}
This result allows us to calculate the time evolution operator to be 
(see Appendix A for details)
\begin{eqnarray}
\tilde{U}(t)&=&\exp\left[-i\int\nolimits_{0}^{t}d\tau\sum_{\lambda}g_{\lambda}\sigma_z\left(e^{i\omega_{\lambda}\tau}a^{\dagger}_{\lambda}+e^{-i\omega_{\lambda}\tau}a_{\lambda}\right)\right]\nonumber\\
&&\times\exp\left(\frac{1}{2}\int_{0}^{t}d\tau\int_{0}^{t}ds\sum_{\lambda}|g_{\lambda}|^2 e^{i\omega_{\lambda}(\tau-s)}\right)\nonumber\\
&&\times\exp\left(-\int_{0}^{t}d\tau\int_{0}^{\tau}ds\sum_{\lambda}|g_{\lambda}|^2 e^{-i\omega_{\lambda}(\tau-s)}\right).
\label{UI}
\end{eqnarray}
The time integrations in 
the exponents in Eq.~(\ref{UI}) can be easily and analytically carried out. But we keep them in those forms in Eq.~(\ref{UI}) 
so it will be easier to identify them with the results in  Ref.~\cite{Goan09}.
If the time-ordering operation in Eq.~(\ref{eq:evolution_op}) 
for $\tilde{U}(t)$ were not performed, one could have just obtained  
the first term (line) of Eq.~(\ref{UI}) for $\tilde{U}(t)$.
Thus the second and third terms (lines) of Eq.~(\ref{UI}) can be considered as
the correction terms due to the time-ordering operation.

The reduced density matrix can be obtained by tracing over the reservoir's degrees of freedom: $\rho(t)={\rm Tr}_R[\rho_T(t)]$.
Suppose initially the state $\rho_{T}(0)=\tilde{\rho}_{T}(0)=\rho(0)\otimes R_0$ is factorized, where $\rho(0)$ and $R_0$ are initial system and thermal reservoir(environment) density operators, respectively, and 
$R_0=\exp(-H_R/k_BT)/{\rm Tr}_R[\exp(-H_R/k_BT)]$.
Then the reduced density matrix elements in the interaction picture can be written as 
\begin{equation}
\tilde{\rho}_{mn}(t)=\rho_{mn}(0){\rm Tr}_R\left(\tilde{U}^{\dagger\{n\}}(t)\tilde{U}^{\{m\}}(t)R_0\right),  
\label{eq:density_elements}
\end{equation}
where $\tilde{\rho}_{mn}(t)\equiv \langle m |\tilde{\rho}(t)|n\rangle$, 
$\tilde{U}^{\{n\}}(t)\equiv\langle n|\tilde{U}(t)|n\rangle$, 
$m,n=0,1$ and the states of the two-level system are defined as
$\sigma_z|0\rangle=|0\rangle$, $\sigma_z|1\rangle=-|1\rangle$.
To evaluate Eq.~(\ref{eq:density_elements}), the well known formula of 
\begin{equation}
e^Ae^B=e^{A+B}e^{\frac{1}{2}[A,B]},  
\label{eq:op_id}
\end{equation}
valid for operators $A$ and $B$ both commuting with the commutator $[A,B]$, can be used to combine the evolution operators together. 
One then obtains 
\begin{eqnarray}
&& \tilde{U}^{\dagger\{0\}}(t_1)\tilde{U}^{\{1\}}(t_1)=
\left[\tilde{U}^{\dagger\{1\}}(t_1)\tilde{U}^{\{0\}}(t_1)\right]^{\dagger} \nonumber \\
&=&
\exp\left[2i\int_{0}^{t_1}d\tau\sum_{\lambda}g_{\lambda}\left(e^{i\omega_{\lambda}\tau}a^{\dagger}_{\lambda}+e^{-i\omega_{\lambda}\tau}a_{\lambda}\right)\right]. 
\label{eq:U0U1}  
\end{eqnarray}
Then a useful identity  \cite{Mermin66}  for the average over the thermal 
reservoir (environment) density operator, $R_0$, can be employed:
\begin{equation}
  \left\langle e^{\sum_{\lambda}c_{\lambda}a_{\lambda}+d_{\lambda}a_{\lambda}^{\dagger}}\right\rangle
=e^{\frac{1}{2}\sum_{\lambda}c_{\lambda}d_{\lambda}(2\bar{n}_{\lambda}+1)},
\label{eq:Mermin_id}
\end{equation}
where $c_\lambda$ , $d_\lambda$ are complex numbers, and $\bar{n}_{\lambda}=[\exp(\hbar\omega_{\lambda}/k_{B}T)-1]^{-1}$
stands for the thermal mean occupation number of the environment oscillators.
As a result, we obtain
\begin{eqnarray}
{\rm Tr}_{R}\left[\tilde{U}^{\dagger\{0\}}(t)\tilde{U}^{\{1\}}(t)R_0\right]
&=&{\rm Tr}_{R}\left[\tilde{U}^{\dagger\{1\}}(t)\tilde{U}^{\{0\}}(t)R_0\right]
\nonumber \\
%&=&e^{-\int_{0}^{t}d\tau D(\tau)},
&=&\exp\left(-\int_{0}^{t}d\tau D(\tau)\right),
\label{TrU0U1}
\end{eqnarray}
where 
\begin{equation}
D(\tau)=2\int_{0}^{\tau}ds[\alpha_{\rm eff}(\tau-s)+\alpha^{\ast}_{\rm eff}(\tau-s)], 
\label{eq:Dt}  
\end{equation}
\begin{equation}
\alpha_{\rm eff}(\tau-s)=\alpha(\tau-s)+\beta(\tau-s), \label{alpha_eff}
\end{equation} 
and
$\alpha(t-\tau)$ and $\beta(t-\tau)$ are defined 
in Eqs.~(\ref{alpha}) and (\ref{beta}), respectively. 
%\begin{eqnarray}
%\alpha_{\rm eff}(\tau-s)&=&\alpha(\tau-s)+\beta(\tau-s), \label{alpha_eff}\\ 
%\alpha(\tau-s)&=&\sum_{\lambda}(\bar{n}_{\lambda}+1)|g_{\lambda}|^2 
%e^{-i\omega_{\lambda}(\tau-s)}, \label{alpha}\\
%\beta(\tau-s)&=&\sum_{\lambda}\bar{n}_{\lambda}|g_{\lambda}|^2 
%e^{i\omega_{\lambda}(\tau-s)}. \label{beta} 
%\end{eqnarray}
%Note that $\alpha(\tau-s)$ and $\beta(\tau-s)$ 
%are also known as the environment CF's:
%$\alpha(\tau-s)=\left\langle
%\sum_{\lambda}g_{\lambda}\tilde{a}_{\lambda}(\tau)\sum_{\lambda'}g_{\lambda'}\tilde{a}^{\dagger}_{\lambda'}(s)\right\rangle$
%and $\beta(\tau-s)=\left\langle
%\sum_{\lambda}g_{\lambda}\tilde{a}^{\dagger}_{\lambda}(\tau)
%\sum_{\lambda'}g_{\lambda'}\tilde{a}_{\lambda'}(s)\right\rangle$,
%where $\tilde{a}_{\lambda}(\tau)=a_{\lambda}e^{-i\omega_\lambda
%\tau}$ and
%$\tilde{a}^{\dagger}_{\lambda}(\tau)=a^{\dagger}_{\lambda}e^{i\omega_\lambda
%\tau}$ are the reservoir operators in the interaction picture. 
%Similarly, we obtain the same result as Eq.~(\ref{TrU0U1}) for 
%${\rm Tr}_{R}\left(\tilde{U}^{\dagger\{0\}}(t)\tilde{U}^{\{1\}}(t)R_0\right)$. 
It is easy to show that $\tilde{U}^{\dagger\{n\}}(t)\tilde{U}^{\{n\}}(t)=I$ and ${\rm Tr}_{R}\left[\tilde{U}^{\dagger\{n\}}(t)\tilde{U}^{\{n\}}(t)R_0\right]=1$. Thus, using these results for the reduced density matrix elements Eq.~(\ref{eq:density_elements}) in the interaction picture and then transforming them back to the Schr\"odinger picture $\rho(t)=\exp\left(-iH_{S}t/\hbar\right)\tilde{\rho}(t)\exp\left(iH_{S}t/\hbar\right)$, we obtain the exact reduced density operator in the matrix form of
\begin{equation}
\rho\left(t\right)=\left(\begin{array}{ccc}\rho_{00}\left(0\right)&\rho_{01}\left(0\right)e^{-F\left(t\right)}\\
\rho_{10}\left(0\right)e^{-F^{\ast}\left(t\right)}&\rho_{11}\left(0\right)\end{array}\right)
\label{density_matrix_solution}
\end{equation}
with $F(t)=i\omega_{S}t+\int_{0}^{t}d\tau D(\tau)$.
The same result was obtained in  Ref.~\cite{Diosi98} using the stochastic Schr\"odinger equation approach. 

With the exact time evolution of the reduced density matrix, the one-time expectation value of the system operators  
\begin{equation}
\left\langle A\left( t_{1}\right) \right\rangle 
={\rm Tr}_{S\oplus R}\left[A\left( t_{1}\right) \rho_{T}\left( 0\right)\right]
={\rm Tr}_{S}\left[A\left( 0\right)\rho\left( t_1\right) \right],  
\label{A1}
\end{equation}
can be calculated exactly, where $A(t_1)$ represents a general system Heisenberg operator(s) and $\rho(t_1)={\rm Tr}_R[\rho_T(t_1)]$ is the reduced Schr\"{o}dinger density matrix operator at time $t_1$. 
%For example, 
We may also write in the interaction picture,
\begin{equation}
\left\langle A\left( t_{1}\right) \right\rangle 
={\rm Tr}_{S\oplus R}\left[\tilde{A}\left( t_{1}\right) \tilde{\rho}_{T}\left( t_1\right)\right]
= {\rm Tr}_{S}\left[\tilde{A}\left( t_{1}\right) 
\tilde{\rho}\left( t_1\right)\right] 
\label{eq:A1I}  
\end{equation}
where $\tilde{\rho}_{T}$ is defined in Eq.~(\ref{eq:rhoTI}), $\tilde{\rho}(t)={\rm Tr}_{R}[\tilde{\rho}_{T}\left(t\right)]$ and $\tilde{A}(t)={\exp}\left(iH_{0}t/\hbar\right)A\,{\exp}\left(-iH_{0}t/\hbar\right)$, and $A=A(0)$. 
For a general system operator 
$A=\left(\begin{array}{ccc}c&a\\b&d\end{array}\right)$,
we obtain  exactly from either Eq.~(\ref{A1}) or Eq.~(\ref{eq:A1I})
\begin{eqnarray}
{\left\langle A\left( t_{1}\right) \right\rangle}
&=&e^{-\int_{0}^{t_1}d\tau D(\tau)}\left(a\rho_{10} e^{i\omega_{S}t_1}+b\rho_{01} e^{-i\omega_{S}t_1}\right)\nonumber \\
&&+c\rho_{00}(0)+d\rho_{11}(0).
\label{exactA1}
\end{eqnarray}

\subsection{Two-time correlation functions}
\label{sec:Exact_2TCF}
 In contrast to the Markovian case in which the QRT is valid, the time evolution of the reduced density matrix of a non-Markovian open system alone
is not sufficient to obtain the two-time system operator CF's. 
This can be understood as follows. The two-time CF's 
of system operators $A(t_1)B(t_2)$ for $t_1>t_2$ can be written as 
\begin{eqnarray}
&&{\left\langle A(t_{1}) B(t_{2})\right\rangle } \notag \\
&=&{\rm Tr}_{S\oplus R}[U^{\dagger}(t_1,0)AU(t_1,0)U^{\dagger}(t_2,0)BU(t_2,0)
\rho_T(0)]
\nonumber\\
&=&{\rm Tr}_{S\oplus R}[AU(t_1,t_2)BU(t_2,0)\rho_T(0)U^{\dagger}(t_2,0)
U^{\dagger}(t_1,t_2)],
\label{2TCF}
\end{eqnarray}
where the Heisenberg evolution operators 
%$U(t,0)=\exp(-i H t/\hbar)$ and 
$U(t_1,t_2)=U(t_1,0)U^{\dagger}(t_2,0)$ 
and $U(t,0)=\exp(-i H t/\hbar)$. 
%We may denote $\rho_T(t_2)=U(t_2,0)\rho_T(0)U^{\dagger}(t_2,0)$. 
If the environment is Markovian so 
the environment operator CF at two different times is 
$\delta$ correlated in time, then we may regard that 
the environment operator in 
$U(t_1,t_2)$ is not correlated with that in $U(t_2,0)$. 
So the trace over the environment degrees of freedom for operator 
$U(t_1,t_2)$ and operator $U(t_2,0)$ can be performed 
independently or separately.
Thus one may first trace $\rho_T(t_2)=U(t_2,0)\rho_T(0)U^{\dagger}(t_2,0)$
over the environment degrees of freedom 
to obtain
the reduced density matrix $\rho(t_2)={\rm Tr}_R[\rho_T(t_2)]$.
Equation (\ref{2TCF}) in this case can be written as   
\begin{eqnarray}
{\left\langle A(t_{1}) B(t_{2})\right\rangle }
&=&{\rm Tr}_{S\oplus R}[AU(t_1,t_2)(B\rho(t_2)\otimes R_0)
U^{\dagger}(t_1,t_2)] \nonumber \\ 
&=&{\rm Tr}_{S}[A\chi(\tau)],
\label{2TCF2}
\end{eqnarray}
where $\chi(\tau)$ is the effective reduced density matrix at time $\tau=t_1-t_2$ with the initial condition $\chi(0)=B\rho(t_2)$. Thus knowing the time evolution of the reduced density matrix in the Markovian case, one is able to calculate the two-time CF's of the system operators. 
This is also the reason why the QRT works in the Markovian case.
But the situation differs for a non-Markovian environment as the environment operator in 
$U(t_1,t_2)$ may, in general, be correlated with that in $U(t_2,0)$.

The two-time CF's of the system operators for the pure-dephasing spin-boson model can also be evaluated exactly. To evaluate 
the two-time CF of system operators  $A(t_1)B(t_2)$ for $t_1>t_2$, 
we express it in terms of the interaction picture operators as   
\begin{eqnarray}
&&{\left\langle A(t_{1}) B(t_{2})\right\rangle } \notag \\
&=&{\rm Tr}_{S\oplus R}[U^{\dagger}(t_1)AU(t_1)U^{\dagger}(t_2)BU(t_2)\rho_T(0)]
\nonumber\\
&=&{\rm Tr}_{S\oplus R}[\tilde{U}^{\dagger}(t_1)\tilde{A}(t_1)\tilde{U}(t_1)\tilde{U}^{\dagger}(t_2)\tilde{B}(t_2)\tilde{U}(t_2)\rho_T(0)],\notag\\
\label{2TCFI}
\end{eqnarray}
where again an operator with a tilde on the top indicates that it is an operator in the interaction picture with respect to the free Hamiltonian $H_{0}$. 
Compared with Eq.~(\ref{eq:A1I}), Eq.~(\ref{2TCFI}) for general non-Markovian open systems can not be expressed as a product of the reduced density matrix and system operators. So again, the reduced density matrix alone is not sufficient to obtain the non-Markovian two-time system operator CF's.

As we want to compare the results by the direct evaluation with those by
the evolution equation (\ref{2time_evol_eq_f}), 
%in Ref.~\cite{Goan09}, 
we calculate, in the following, the two-time CF's $\langle A(t_{1}) B(t_{2})\rangle$ for different cases of system operators $A$ and $B$.
The structure of the evolution equations in Ref.~\cite{Goan09} or Eqs.~(\ref{1time_evol_eq_f}) and (\ref{2time_evol_eq_f}) in this article depends on the commutation relations of operator $A$ and operator $L$ (or $L^\dagger$), and on the commutation relations of operator $B$ and operator $\tilde{L}(\tau-t_2)$ 
(or $\tilde{L}^\dagger(\tau-t_2)$), where $\tilde{L}(t)=\exp\left(iH_St/\hbar\right)L\exp\left(-iH_St/\hbar\right)$ is the system operator in the interaction picture with respect to $H_S$.
For the pure-dephasing spin-boson model, $H_S=(\hbar\omega_S/2)\sigma_z$, $L=\sigma_z=L^{\dagger}$, and then $\tilde{L}^{\dagger}(t)=\sigma_z$. 
So we will discuss the two-time CF's in the following three cases
and the trivial case of $\langle \sigma_z(t_{1})\sigma_z(t_{2})\rangle=\langle\sigma_z(0)\sigma_z(0)\rangle=1$ is obvious due to $[\sigma_z,H]=0$. 

{\em{Case 1}.} $[A,L]\neq 0$ and $[B,\tilde{L}(t)]=0$.
%$A=\left(\begin{array}{ccc}0&a\\b&0\end{array}\right)
 In this case, let us set 
$A=a\sigma_{+}+b\sigma_{-}$, and $B=\sigma_z$. Then 
$\tilde{A}(t)=a\sigma_{+}e^{i\omega_{S}t}+b\sigma_{-}e^{-i\omega_{S}t}$ and 
$\tilde{B}(t)=\sigma_z$.  It is easy to see from Eq.(\ref{UI}) that 
$\tilde{U}(t)$ commutes with $\tilde{B}(t)$ but anticommutes with 
$\tilde{A}(t)$, 
i.e., $[\tilde{U}(t),\tilde{B}(t)]=0$ and $\{\tilde{U}(t),\tilde{A}(t)\}=0$. 
Using these results and the fact that $U^{\dagger \{n\}}(t)U^{\{n\}}(t)=I$, 
we obtain from Eq.~(\ref{2TCFI}) 
\begin{eqnarray}
&&{\left\langle A(t_{1})B(t_{2})\right\rangle}\notag\\
&=&-a\rho_{10}(0) e^{i\omega_{S} t_{1}}{\rm Tr}_{R}[\tilde{U}^{\dagger \{0\}}(t_1)\tilde{U}^{\{1\}}(t_1)R_0]\nonumber\\
&&+b\rho_{01}(0) e^{-i\omega_{S} t_{1}}{\rm Tr}_{R}[\tilde{U}^{\dagger \{1\}}(t_1)\tilde{U}^{\{0\}}(t_1)R_0].
\label{CF1}
\end{eqnarray}  
Substituting the result of Eq.~(\ref{TrU0U1}) into Eq.~(\ref{CF1}),
we arrive at the exact two-time CF's 
\begin{eqnarray}
&&{\left\langle A(t_{1})B(t_{2})\right\rangle}\notag\\
&=&e^{-\int_{0}^{t_1}d\tau D(\tau)}\left(-a\rho_{10}(0) e^{i\omega_{S} t_{1}}+b\rho_{01}(0) e^{-i\omega_{S} t_{1}}\right).\notag\\
\label{CF1result}
\end{eqnarray}
%If we choose $a=b=1$ ( i.e. A= $\sigma_{x}$) and $a=-b=-i$ 
%( i.e. A= $\sigma_{y}$), and then take the derivative of Eq.~(\ref{CF1result})
%, we can obtain the same results as Eqs.~(\ref{2time_evol_xz}) 
%and (\ref{2time_evol_yz}), respectively.\\

{\em{Case 2}.} $[A,L]=0$ and $[B,\tilde{L}(t)]\neq 0$.
In this case, let $A=\sigma_z$, and $B=a\sigma_{+}+b\sigma_{-}$. Similar to the calculations in Case 1, we obtain
%\begin{eqnarray}
%&&{\left\langle A(t_{1})B(t_{2})\right\rangle}\notag\\
%&=&a\rho_{10}(0) e^{i\omega_{S} t_{2}}
%{\rm Tr}_{R}\left(U^{\dagger \{0\}}_I(t_2)U^{\{1\}}_I(t_2)R_0\right)\notag\\
%&&-b\rho_{01}(0) e^{-i\omega_{S} t_{2}}
%{\rm Tr}_{R}\left(U^{\dagger \{1\}}_I(t_2)U^{\{0\}}_I(t_2)R_0\right) ,
%\label{CF3}
%\end{eqnarray}
%Using the result in Eq.~(\ref{TrU0U1}) but replacing $t_1$ by $t_2$, we have
\begin{eqnarray}
&&{\left\langle A(t_{1})B(t_{2})\right\rangle}\notag\\
&=&e^{-\int_{0}^{t_2}d\tau D(\tau)}\left(a\rho_{10}(0) e^{i\omega_{S} t_{2}}-b\rho_{01}(0) e^{-i\omega_{S} t_{2}}\right).\notag\\
\label{CF2result}
\end{eqnarray}
%One can investigate that the CF in Eq.~(\ref{CF3result}) 
%is independent of $t_1$, thus its evolution with $t_1$ is consistent with 
%Eqs.~(\ref{2time_evol_zx}) and (\ref{2time_evol_zy}).\\
The exact two-time CF's of Eqs.~(\ref{CF1result}) and (\ref{CF2result}) depend on only one time variable, $t_1$ or $t_2$, respectively, since one of the system operator $\sigma_z(t)=\sigma_z(0)$ is time-independent.

{\em{Case 3}.} $[A,L] \neq 0$ and $[B,\tilde{L}(t)]\neq 0$.
Suppose $A=a\sigma_{+}+b\sigma_{-}$, and
$B=a^{\prime}\sigma_{+}+b^{\prime}\sigma_{-}$. 
In this case, both $\tilde{A}(t)$ and $\tilde{B}(t)$ anticommute with 
both $\tilde{U}(t)$
and $\tilde{U}^{\dagger}(t)$.
%$\left\{\tilde{A}(t),\tilde{U}(t)\right\}=0$ and 
%$\left\{\tilde{B}(t),\tilde{U}(t)\right\}=0$.
Furthermore, $\tilde{A}(t_1)\tilde{B}(t_2)=ab^{\prime}\sigma_{+}\sigma_{-}\exp[i\omega_{S}(t_1-t_2)] + ba^{\prime}\sigma_{-}\sigma_{+}\exp[-i\omega_{S}(t_1-t_2)]$.
%Note that in this case, $\left\{B_I(t),U_I(t)\right\}=0$. 
Thus we can obtain from Eq.~(\ref{2TCFI}) 
\begin{eqnarray}
&&{\left\langle A(t_{1})B(t_{2})\right\rangle}\notag\\
&=&ab^{\prime}\rho_{00}(0)e^{i\omega_{S}(t_1-t_2)}\notag\\
&&\times{\rm Tr}_R[\tilde{U}^{\dagger\{0\}}(t_1)\tilde{U}^{\{1\}}(t_1)\tilde{U}^{\dagger\{1\}}(t_2)\tilde{U}^{\{0\}}(t_2)R_0]\notag\\
&&+ ba^{\prime}\rho_{11}(0)e^{-i\omega_{S}(t_1-t_2)}\notag\\
&&\hspace{-0.2cm}\times{\rm Tr}_R[\tilde{U}^{\dagger\{1\}}(t_1)\tilde{U}^{\{0\}}(t_1)\tilde{U}^{\dagger\{0\}}(t_2)\tilde{U}^{\{1\}}(t_2)R_0].
\label{CF2}
\end{eqnarray}
%Since $U^{\dagger\{0\}}_I(t_1)U^{\{1\}}_I(t_1)=\left(U^{\dagger\{1\}}_I(t_2)U^{\{0\}}_I(t_2)\right)^{\dagger}=
%\exp\left(2i\int_{0}^{t_1}d\tau\sum_{\lambda}g_{\lambda}\left(e^{i\omega_{\lambda}\tau}a^{\dagger}_{\lambda}+e^{-i\omega_{\lambda}\tau}a_{\lambda}\right)\right)$
It is obvious from Eq.~(\ref{CF2}) that to evaluate the general two-time CF, we need to take into account the correlations of the reservoir operators 
of the evolution operators between different time periods of $[0,t_2]$ and $[0,t_1]$ before the trace over the environment is performed. 
Using Eqs.~(\ref{eq:op_id}), (\ref{eq:U0U1}),  and (\ref{eq:Mermin_id}),
 we get 
\begin{eqnarray}
&&{\rm Tr}_R[\tilde{U}^{\dagger\{0\}}(t_1)\tilde{U}^{\{1\}}(t_1)\tilde{U}^{\dagger\{1\}}(t_2)\tilde{U}^{\{0\}}_I(t_2)R_0]\notag\\
&=&\exp\left[
%-\left(\int_{0}^{t_1}d\tau+\int_{0}^{t_2}d\tau\right)D(\tau)
-\int_{0}^{t_1}d\tau D(\tau)-\int_{0}^{t_2}d\tau D(\tau)
+\int_{0}^{t_1}d\tau\tilde{D}(\tau,t_2)\right],\notag\\
\label{TrU0U1U1U0}
\end{eqnarray}  
where 
\begin{equation}
\tilde{D}(\tau,t_2)=4\int_{0}^{t_2}ds\, \alpha_{\rm eff}(\tau-s).
\label{eq:Dt1t2}  
\end{equation}
The term $\int_{0}^{t_1}d\tau\tilde{D}(\tau,t_2)$ in Eq.~(\ref{TrU0U1U1U0})
describes the cross-time contribution of 
the environment CF's 
of the reservoir operators in 
the evolution operators $\tilde{U}^{\{n\}}(t_1)$ and $\tilde{U}^{\dagger\{n\}}(t_2)$ 
[or $\tilde{U}^{\dagger\{n\}}(t_1)$ and $\tilde{U}^{\{n\}}(t_2)$] of 
the two different time periods $[0,t_1]$ and $[0,t_2]$. 
We can see this from 
$\tilde{D}(\tau,t_2)$ of Eq.~(\ref{eq:Dt1t2}) and 
in Eq.~(\ref{TrU0U1U1U0}) that
the environment CF $\alpha_{\rm eff}(\tau-s)$, defined 
in Eq.~(\ref{alpha_eff}), has the time variable $\tau$ in $[0,t_1]$
and the time variable $s$ in $[0,t_2]$.
%and $\alpha_{\rm eff}(\tau-s)$ is defined in Eq.~(\ref{alpha_eff}). 
%where we have used the well known formula of 
%$\exp(A+B)=\exp(A)\exp(B)\exp\left(-\frac{1}{2}[A,B]\right)$ 
%conditioned on $A$ and $B$ are both commuted with the commutator $[A,B]$. 
On the other hand, the time evolution of the reduced density matrix (\ref{eq:density_elements}) 
is involved with the reservoir operator CF's in the evolution operators of 
only the same time interval. 
As a result, it, alone, cannot provide us with the full information 
to evaluate the non-Markovian two-time CF, 
even in the weak system-environment coupling case.
Similarly, we find that ${\rm Tr}_R[\tilde{U}^{\dagger\{1\}}_I(t_1)\tilde{U}^{\{0\}}_I(t_1)\tilde{U}^{\dagger\{0\}}_I(t_2)\tilde{U}^{\{1\}}_I(t_2)R_0]$ has the same result as Eq.~(\ref{TrU0U1U1U0}).
Substituting these results into Eq.~(\ref{CF2}), 
finally we arrive at the two-time CF
\begin{eqnarray}
&&{\left\langle A(t_{1})B(t_{2})\right\rangle}\notag\\
\hspace{-0.3cm}&=&
%\exp\left[-\left(\int_{0}^{t_1}d\tau+\int_{0}^{t_2}d\tau\right)D(\tau)
\exp\left[-\int_{0}^{t_1}d\tau D(\tau)-\int_{0}^{t_2}d\tau D(\tau)
+\int_{0}^{t_1}d\tau\tilde{D}(\tau,t_2)\right]\notag\\
&&\hspace{-0.4cm}\times\left(ab^{\prime}\rho_{00}(0)e^{i\omega_{S}(t_1-t_2)}+ba^{\prime}\rho_{11}(0)e^{-i\omega_{S}(t_1-t_2)}\right).
\label{CF3result}
\end{eqnarray}
This non-Markovian finite-temperature two-time CF, to our knowledge, 
has not been presented in the literature. 
%Let $a$($a^{\prime}$)'s and $b$($b^{\prime}$)'s be equal to $\pm 1$ and $\pm i$, and then taking the derivative of Eq.~(\ref{CF2result})
%with respect to $t_1$, we arrive the same evolution equations as Eqs.~(\ref{2time_evol_xy}) to (\ref{2time_evol_yy}).\\

\section{Evaluation by derived non-Markovian finite-temperature evolution equations} 
\label{derived_EE}
In this section, we will use the derived evolution equations in Ref.~\cite{Goan09} to compute the one-time expectation values and two-time CF's 
to compare with the exact expressions evaluated in Sec.~\ref{sec:Exact}.
Despite the fact that the evolution equations in Ref.~\cite{Goan09}
derived perturbatively, the results obtained this way for the
pure-dephasing spin-boson model happen to be the same as the exact
expressions by the direct evaluation. 

\subsection{Quantum master equation and one-time expectation values}
Before going to calculate the CF's, it is instructive to derive the
master equation of the reduced system density matrix for the model. 
%Since we consider 
%only up to second order in the system-environment interaction strength, 
%we can replace $\tilde{\rho}_T(t')$ in Eq.~(5) of Ref.~\cite{Goan09} by 
%$\tilde{\rho}_T(t)=\tilde{\rho}(t)\otimes R_0+ \cal{O}(\tilde{H}_I)$ 
%\cite{Goan09}. 
%and at all time the total density operator should only show derivations 
%of order $H_I$ from an uncorrelated (factorized) state, 
%$\tilde{\rho}_T(t)=\tilde{\rho}(t)\otimes R_0+O(H_I)$\cite{Carmichael99}. 
%In our work, we only take the validity to the second order in the interaction 
%Hamiltonian, hence we can approximate 
%the total density operator to a factorized form: $\tilde{\rho}_T(t)\approx\tilde{\rho}(t)\otimes R_0$ in Eq.~(5) of Ref.~\cite{Goan09}.
%Tracing over the environment degrees of freedom for the resultant equation
%Eq.~(5) in Ref.~\cite{Goan09} 
%and then transferring from the interaction picture back to the Schr\"odinger through $\rho(t)=\exp(-iH_St/\hbar)\tilde{\rho}(t)\exp(iH_St/\hbar)$, 
After some calculations, 
we obtain for the Hamiltonian in the form of Eq.~(\ref{BosonModel}) a
time-covolutionless non-Markovian 
master equation \cite{Breuer02,Shibata77, Breuer99, Schroder06, Ferraro09,Paz01,Goan07}
valid to second order in the system-environment interaction
strength 
\begin{eqnarray}
\frac{d \rho(t)}{dt} 
&=&\frac{1}{i\hbar}\left[H_S,\rho(t)\right] 
\notag \\
&&
\hspace{-1cm}
-\int_0^t d\tau \{\alpha(t-\tau)[L^\dagger\tilde{L}(\tau-t)\rho(t)
-\tilde{L}(\tau-t)\rho(t)L^\dagger] \notag\\
&&
\hspace{-0.5cm}
 +\beta(t-\tau)[L\tilde{L}^\dagger(\tau-t)\rho(t)
-\tilde{L}^\dagger(\tau-t)\rho(t)L] \notag \\
&&\hspace{-0.5cm} +{\rm H.c.} \},
\label{master_eq}
%e^{iH_st/\hbar}{\rm Tr}_R\int_{0}^{t}dt^{\prime }
%\left[ \tilde{H}_{I}(t) ,\left[ \tilde{H}_{I}
%( t^{\prime }) ,\tilde{\rho}_{T}( t) \right] \right] e^{-iH_st/\hbar},  
\end{eqnarray} 
%where we have used Hamiltonian in the form of Eq.~(\ref{BosonModel}), 
where $\alpha(t-\tau)$ and $\beta(t-\tau)$ are defined in
Eqs.~(\ref{alpha}) and (\ref{beta}) respectively, H.c. indicates the
Hermitian conjugate of previous terms, and an operator with a tilde on
the top indicates that it is an operator in the interaction picture. 
For the pure-dephasing spin-boson model, Eq.~(\ref{master_eq}) gives the 
master equation of the reduced system density matrix
\begin{equation}
\frac{d\rho\left(t\right)}{dt}=\frac{-i\omega_{S}}{2}\left[\sigma_{z},\rho\left(t\right)\right]
-\frac{D(t)}{2}\left[\rho(t)-\sigma_{z}\rho\left(t\right)\sigma_{z}\right],
\label{PerturbationME}
\end{equation}
where $D(t)$ is defined in Eq.~(\ref{eq:Dt}). 
It is not difficult to show that the exact expression of the density matrix  
(\ref{density_matrix_solution}) is the solution of the master equation 
(\ref{PerturbationME}) although the master equation is derived perturbatively. 
Non-Markovian dynamics usually means that the current time evolution of the system state depends on its history, and the memory effects typically enters through integrals over the past state history. However, the non-Markovian system dynamics of some class of open quantum system models may be summed up and expressed as a time-local, convolutionless form \cite{Strunz04}
where the dynamics is determined by the system state at the current time $t$ only.  This time-local, convolutionless class of open quantum systems may be treated exactly without any approximation. The quantum Brownian motion model or the damped harmonic oscillator bilinearly coupled to 
a bosonic bath of 
harmonic oscillators \cite{Strunz04,Haake86,Hu92} is a famous example of this class. 
The pure-dephasing spin-boson model considered here also belongs to this class, and the non-Markovian effect in the master equation (\ref{PerturbationME}) is taken into account by the time-dependent coefficient $D(t)$ 
instead of memory integral. 
This time-local, convolutionless property and 
the fact of $[L, H_s]=0$ allow the exact system density matrix 
Eq.~(\ref{density_matrix_solution}) to be obtained 
from Eq.~(\ref{PerturbationME}). 
%that replaces the memory integrals usually encountered 
%in other open quantum system models.

Since the exact solution of the system density matrix
(\ref{density_matrix_solution}) can be calculated from the
perturbatively derived master equation (\ref{PerturbationME}), one may
expect that the exact non-Markovian finite-temperature one-time
expectation values and two-time CF's of the pure-dephasing model can
be obtained from the evolution equation (\ref{2time_evol_eq_f}) 
%derived in Ref.~\cite{Goan09}.  
We show below that this is indeed the
case, and at the same time the agreement of the results demonstrates
the validity and practical usage of the evolution equation
(\ref{2time_evol_eq_f}).
% in Ref.~\cite{Goan09}.  

For the pure-dephasing spin-boson model, $H_S=(\hbar\omega_S/2)\sigma_z$, $L=\sigma_z=L^{\dagger}$, and we have $\tilde{L}^{\{\dagger\}}(t)=\sigma_z$. 
Taking $A=\sigma_{i}$, $i=x,y,z$, in Eq.~(\ref{1time_evol_eq_f}), we obtain
straightforwardly 
the evolution equations of the single-time expectation values as
\begin{eqnarray}
d\left\langle \sigma_x(t_{1})\right\rangle/dt_{1}&=&-D(t_{1})\left\langle \sigma_x(t_{1})\right\rangle-\omega_{S}\left\langle \sigma_y(t_{1})\right\rangle,
\label{1time_evol_x}\\ 
 d\left\langle \sigma_y(t_{1})\right\rangle/dt_{1}&=&-D(t_{1})\left\langle \sigma_y(t_{1})\right\rangle+\omega_{S}\left\langle \sigma_x(t_{1})\right\rangle,\
\label{1time_evol_y}\\
d\left\langle \sigma_z(t_{1})\right\rangle/dt_{1}&=&0
\label{1time_evol_z}
\end{eqnarray}
with $D(t_1)$ defined in Eq.~(\ref{eq:Dt}).  
With proper chosen values for $a$, $b$, $c$, and $d$ of a general operator 
$A$ for $\sigma_i$, 
one can verify that the exact expression of the expectation value of 
$\sigma_i(t_1)$ in Eq.~(\ref{exactA1}) satisfies 
Eqs.~(\ref{1time_evol_x})--(\ref{1time_evol_z}).

\subsection{Two-time correlation functions}
Before using Eq.~(\ref{2time_evol_eq_f}) to calculate the non-Markovian finite-temperature two-time CF's, we discuss briefly below the relation between the QRT and the evolution equation (\ref{2time_evol_eq_f}). 
If the last two terms of Eq.~(\ref{2time_evol_eq_f}) vanish, then the single-time and two-time evolution equations (\ref{1time_evol_eq_f}) and (\ref{2time_evol_eq_f}) will have the same form with the same evolution coefficients and thus the QRT will be applicable. 
The last two terms of Eq.~(\ref{2time_evol_eq_f}) or more generally 
the last term of Eq.~(17) in Ref.\cite{Goan09} 
involve(s) the propagation from $\tau=0$ to $\tau=t_{2}$,
and these terms would vanish for the 
CF's $\langle A(t_1)B(0)\rangle$ as $t_2=0$ in this case.
So the QRT is valid to 
calculate the CF's $\langle A(t)B(0)\rangle$ of both Markovian and  non-Markovian open systems, where the system-environment density matrix is 
separable at $t=0$. 
%But the QRT is also valid and actually applied to calculate 
The QRT is also valid and is often applied to calculate, 
in the Markovian weak system-environment coupling case,
more general CF's $\langle A(t_2+\tau) B(t_2)\rangle$ or equivalently 
$\langle A(t_1)B(t_2)\rangle$ with $t_2 \neq 0$. 
For example, the QRT is used to calculate the Markovian steady-state CF's, and in this case $t_2$ is set to any of the large times when the steady state
%, in which the state and expectation values do not change with time any more, 
is reached.   
This is because in the Markovian case, the last two terms of Eq.~(\ref{2time_evol_eq_f}) vanish since the time integration of the corresponding $\delta$-correlated reservoir CF's, $\alpha(t_1-\tau)\propto \delta(t_1-\tau)$ and $\beta(t_1-\tau)\propto \delta(t_1-\tau)$, over the variable $\tau$ in the domain from $0$ to $t_2$ is zero as $t_1>t_2$.
%In general, $\langle A(t_1)B(t_2)\rangle$ depends on the values of 
%$t_1$ and $t_2$. 
%The initial condition $\langle A(t_2) B(t_2)\rangle$ for 
%the two-time evolution equation (\ref{2time_evol_eq_f}) 
%(without the last two terms for the QRT)  
%can be calculated from the one-time evolution equation 
%(\ref{1time_evol_eq_f}). 
%Or alternatively, the initial effective reduced density operator 
%$\chi(0)=B \rho(t_2)$ for (\ref{2TCF2}) can be calculated using 
%Eq.~(\ref{PerturbationME}).  
%But once the ``initial'' condition is known, 
%$\langle A(t_2) B(t_2)\rangle$ or $\chi(0)=B \rho(t_2)$ at $t_2$ is known, 
%then in the Markovian QRT case, the further CF evolution depends only 
%on $t_1$, or more precisely only on the time difference $(t_1-t_2)$.   
On the other hand, the QRT cannot be blindly applied to calculate $\langle A(t_1)B(t_2)\rangle$ with $t_2 \neq 0$ in a general non-Markovian open system 
%due to the presence of the cross CF's of the reservoir operators 
%at two different times:
%On the other hand, to calculate $\langle A(t_1)B(t_2)\rangle$ with 
%$t_2 \neq 0$ in a general non-Markovian open system, the QRT does not work
%the further CF evolution not only depends on $t_1$ but also 
%depends on $t_2$. [see the last two terms in Eq.~(\ref{2time_evol_eq_f})]. 
due to the non-vanishing contributions of the cross correlation of the reservoir operators at two different times: a later time $t_1$ and an earlier time in the period between $0$ and $t_2$ (see the last two terms of Eq.~(\ref{2time_evol_eq_f}) and also Fig.~\ref{fig:CF_t2_02}).  In other words, in contrast to the Markovian case, not only the initial condition $\langle A(t_2) B(t_2)\rangle$ for the two-time evolution equation (\ref{2time_evol_eq_f}) but also the equation (\ref{2time_evol_eq_f}) itself may depend on the choice of the starting time $t_2$ of the non-Markovian finite-temperature two-time CF's. In the steady state, the situation may change when $t_2$ is in any of the large times where the state and system expectation values do not change with time any more. In this case, the contributions from the last two terms of Eq.~(\ref{2time_evol_eq_f}) saturate and do not depend on where time $t_2$ is set in the steady state, and thus both the Markovian and non-Markovian CF's may depend only on the time difference $(t_1-t_2)$ (see also Fig.~\ref{fig:CF_diff_t2}).
But the nonvanishing contributions from the last two terms of Eq.~(\ref{2time_evol_eq_f}) would still make the non-Markovian CF's deviate from that obtained wrongly using the QRT in the non-Markovian case or obtained using the QRT in the Markovian case (see also Fig.~\ref{fig:CF_t2_10}).

For the time evolutions of system two-time CF's of the pure-dephasing 
spin-boson model, we also consider the 
following three cases as in Sec.~\ref{sec:Exact}. 
Note that $\tilde{L}(t)=\sigma_z=\tilde{L}^{\dagger}(t)$.

{\em{Case 1}.} $[A,L]\neq 0$ and $[B,\tilde{L}(t)]=0$.
In this case, let $A=\sigma_{i},i=x,y$ and $B=\sigma_z$. 
%Since $[B,\tilde{L}(t)]=0$ and 
%$[B,\tilde{L}^\dagger(t)]=0$, 
By using Eq.~(\ref{2time_evol_eq_f}), it is easy to obtain
\begin{eqnarray}
d\langle \sigma_x(t_{1}) \sigma_z(t_{2})\rangle/dt_{1}&=&-D(t_{1})\langle \sigma_x(t_{1})\sigma_z(t_{2})\rangle\nonumber\\
&&-\omega_{S}\langle \sigma_y(t_{1}) \sigma_z(t_{2})\rangle,
\label{2time_evol_xz}\\ 
d\langle \sigma_y(t_{1}) \sigma_z(t_{2})\rangle/dt_{1}&=&-D(t_{1})\langle \sigma_y(t_{1})\sigma_z(t_{2})\rangle\nonumber\\
&&+\omega_{S}\langle \sigma_x(t_{1}) \sigma_z(t_{2})\rangle.
\label{2time_evol_yz}
%\\
%d\langle \sigma_z(t_{1}) \sigma_z(t_{2})\rangle/dt_{1}&=&0 ,
\label{2time_evol_zz} 
\end{eqnarray}
In this case, one can see that the evolution equations of single-time
expectation values $\langle\sigma_i(t_1)\rangle$, Eqs.~(\ref{1time_evol_x}) and (\ref{1time_evol_y}), have the same forms as the  
evolution equations of two-time CF's $\langle\sigma_i(t_1)\sigma_z(t_2)\rangle$, Eqs.~(\ref{2time_evol_xz}) and (\ref{2time_evol_yz}), respectively.
Hence the QRT is valid in this case.
It is easy to check that taking the derivative of Eq.~(\ref{CF1result}) with respect to $t_1$ with  
$a=b=1$ ( i.e. A= $\sigma_{x}$) and $a=-b=-i$ ( i.e. A= $\sigma_{y}$), 
one can obtain the evolution equations for 
$\langle\sigma_x(t_1)\sigma_z(t_2)\rangle$ and 
$\langle\sigma_y(t_1)\sigma_z(t_2)\rangle$, exactly the
same as Eqs.~(\ref{2time_evol_xz}) and (\ref{2time_evol_yz}), respectively.

{\em{Case 2}.} $[A,L]=0$ and $[B,\tilde{L}(t)]\neq 0$.
In this case, let $A=\sigma_z$ and  $B=\sigma_{i},i=x,y$. 
%Since $[B,\tilde{L}(t)]=0$ and 
%$[B,\tilde{L}^\dagger(t)]=0$, 
By using Eq.~(\ref{2time_evol_eq_f}), we then easily obtain
\begin{eqnarray}
d\langle \sigma_z(t_{1}) \sigma_x(t_{2})\rangle/dt_{1}&=&0,
\label{2time_evol_zx}\\
d\langle \sigma_z(t_{1}) \sigma_y(t_{2})\rangle/dt_{1}&=&0.
\label{2time_evol_zy}
\end{eqnarray}
Indeed, Eq.~(\ref{CF2result}) satisfies Eqs.~(\ref{2time_evol_zx}) and (\ref{2time_evol_zy}), and 
$\langle\sigma_z(t_1)\sigma_i(t_2)\rangle
=\langle\sigma_z(t_2)\sigma_i(t_2)\rangle$, independent of $t_1$.

{\em{Case 3}.} $[A,\tilde{L}(t)] \neq 0$ and $[B,\tilde{L}(t)]\neq 0$.
In this case, let $A=\sigma_{i},i=x,y$ and  $B=\sigma_{j},j=x,y$. 
Eq.~(\ref{2time_evol_eq_f}) straightforwardly yields
\begin{eqnarray}
d\langle \sigma_x(t_{1}) \sigma_y(t_{2})\rangle/dt_{1}&=&-D(t_{1})\langle \sigma_x(t_{1})\sigma_y(t_{2})\rangle\nonumber\\ 
&&\hspace{-3cm}-\omega_{S}\langle \sigma_y(t_{1}) \sigma_y(t_{2})\rangle-\tilde{D}(t_1,t_{2})\langle \sigma_y(t_{1})\sigma_x(t_{2})\rangle,
\label{2time_evol_xy}\\ 
d\langle \sigma_y(t_{1}) \sigma_x(t_{2})\rangle/dt_{1}&=&-D(t_{1})\langle \sigma_y(t_{1})\sigma_x(t_{2})\rangle\nonumber\\
&&\hspace{-3cm}+\omega_{S}\langle \sigma_x(t_{1}) \sigma_x(t_{2})\rangle-\tilde{D}(t_1,t_{2})\langle \sigma_x(t_{1})\sigma_y(t_{2})\rangle,
\label{2time_evol_yx}\\ 
d\langle \sigma_x(t_{1}) \sigma_x(t_{2})\rangle/dt_{1}&=&-D(t_{1})\langle \sigma_x(t_{1})\sigma_x(t_{2})\rangle\nonumber\\
&&\hspace{-3cm}-\omega_{S}\langle \sigma_y(t_{1}) \sigma_x(t_{2})\rangle+\tilde{D}(t_1,t_{2})\langle \sigma_y(t_{1})\sigma_y(t_{2})\rangle,
\label{2time_evol_xx}\\ 
d\langle \sigma_y(t_{1}) \sigma_y(t_{2})\rangle/dt_{1}&=&-D(t_{1})\langle \sigma_y(t_{1})\sigma_y(t_{2})\rangle\nonumber\\
&&\hspace{-3cm}+\omega_{S}\langle \sigma_x(t_{1}) \sigma_y(t_{2})\rangle+\tilde{D}(t_1,t_{2})\langle \sigma_x(t_{1})\sigma_x(t_{2})\rangle,
\label{2time_evol_yy}
\end{eqnarray}
where $\tilde{D}(t_1,t_2)$ is defined in Eq.~(\ref{eq:Dt1t2}). 
The evolution equations, Eqs.~(\ref{2time_evol_xy})--(\ref{2time_evol_yy}), have different forms as those of single-time expectation values due to the existence of $\tilde{D}(t_1,t_2)$ terms. As a result, the QRT does not hold in this case.
Again, taking the derivative of Eq.~(\ref{CF3result})
with respect to $t_1$ with properly chosen values for $a$, $b$, $a'$ and $b'$, we arrive at the same evolution equations as those from Eqs.~(\ref{2time_evol_xy}) to (\ref{2time_evol_yy}).
Alternatively, solving the coupled equations, Eqs.~(\ref{2time_evol_xy})--(\ref{2time_evol_yy}), one would 
obtain the solutions in a form as Eq.~(\ref{CF3result}).

The agreement between the results obtained by the direct operator evaluation and those obtained by solving the coupled evolution equations demonstrates clearly the validity of the equations (\ref{1time_evol_eq_f}) and 
(\ref{2time_evol_eq_f}), 
%or more generally Eqs.~(14) and (17) in Ref.~\cite{Goan09}. 
In addition, the easiness to obtain Eqs.~(\ref{1time_evol_x})--(\ref{1time_evol_z}) from the evolution equation (\ref{1time_evol_eq_f}), 
and to obtain Eqs.~(\ref{2time_evol_xz})--(\ref{2time_evol_yz}), Eqs.~(\ref{2time_evol_zx}) and (\ref{2time_evol_zy}), and Eqs.~(\ref{2time_evol_xy})--(\ref{2time_evol_yy}) from the evolution equation (\ref{2time_evol_eq_f}) illustrates the practical usage of the non-Markovian finite-temperature evolution equations (\ref{1time_evol_eq_f}) and (\ref{2time_evol_eq_f}).

\section{Results and discussions}
\label{sec:result}
To calculate the two important functions $D(t)$ and $\tilde{D}(t_1,t_2)$, we need to evaluate the environment CF 
\begin{eqnarray}
 \alpha_{\rm eff}(t_1-\tau)
&=&\int_{0}^{\infty}d\omega J(\omega)
\left\{\coth\left({\hbar\omega}/{2k_{B}T}\right)\cos[\omega(t_1-\tau)]\right.
\notag\\
&& \hspace{1.5cm}\left.-i\sin[\omega(t_1-\tau)]\right\}. 
\label{eq:bathCF}  
\end{eqnarray}
where $J(\omega)=\sum_\lambda|g_\lambda|^2\delta(\omega-\omega_\lambda)$ is the spectral density of the environment.
We may consider any spectral density to characterize the environment, but for simplicity we consider an ohmic bath with exponential cut-off function as 
\begin{equation}
J(\omega)=\gamma\omega\exp(-\omega/\Lambda),
\label{Jw}
\end{equation} 
where $\Lambda$ is the cut-off frequency and $\gamma$ is a dimensionless constant characterizing the interaction strength to the environment.
At the zero temperature, the function $D(t)$ and $\tilde{D}(t_1,t_2)$ have simple analytical forms:
%Consequently, we can evaluate coefficients $D(t_1)$ and $\bar{D}(t_2)$ 
%appear in evolution equations of single-time expectation values and 
%two-time CF's.
%For the zero-temperature case, we can obtain these coefficients as
\begin{eqnarray}
D(t_1)
%&=&4\gamma\int_{0}^{t_1}d\tau\int_{0}^{\infty}d\omega\omega 
%e^{-\omega/\Lambda}\cos\left(\omega(t_1-\tau)\right)\notag\\
&=&4\gamma\frac{\Lambda^{2}t_{1}}{1+\Lambda^{2}t_{1}^{2}},
\label{Dt1}\\
%\end{eqnarray}
%and
%\begin{eqnarray}
\tilde{D}(t_1,t_2)
%&=&4\gamma\int_{0}^{t_2}d\tau\int_{0}^{\infty}d\omega\omega 
%e^{-\omega/\Lambda}e^{-i\omega(t_1-\tau)}\notag\\
&=&\frac{4\gamma\Lambda^{2}t_2\left[1-\Lambda^{2}t_1(t_1-t_2)-{i}\Lambda(2t_1-t_2)\right]}{(1+\Lambda^{2}t_1^{2})\left[1+\Lambda^{2}(t_1-t_2)^{2}\right]}.
\label{Dt1t2}
\end{eqnarray}
Consequently, the one-time expectation values and the two-time CF's also have simple analytical expressions. For example, the zero-temperature two-time CF's of Eq.~(\ref{CF1result}) in case 1 and Eq.~(\ref{CF3result}) in case 3 are
\begin{eqnarray}
&&{\left\langle A(t_1)B(t_2)\right\rangle}\notag\\
&=&
%e^{-2\gamma \ln(1+\Lambda^{2}t_{1}^{2})}
{(1+\Lambda^{2}t_{1}^{2})^{-2\gamma}}
\left(-a\rho_{10}(0)e^{i\omega_{S}t_1}+b\rho_{01}(0)e^{-i\omega_{S}t_1}\right)
\label{two_time CF1}
\end{eqnarray} 
and
\begin{eqnarray}
&&{\left\langle A(t_1)B(t_2)\right\rangle}\notag\\
&=&
%\exp\left\{-2\gamma \ln\left[1+\Lambda^{2}(t_{1}-t_{2})^{2}\right]\right.
%\notag\\
[1+\Lambda^{2}(t_{1}-t_2)^{2}]^{-2\gamma}
\notag\\ &&\times
e^{-4\gamma{i}\left(\arctan\left(\Lambda(t_1-t_2)\right)+\arctan\left(\Lambda t_2\right)-\frac{\Lambda t_1}{1+\Lambda^{2}t_{2}^{2}}\right)}
\notag\\
%&&\left.-4\gamma{i}\left(\arctan\left(\Lambda(t_1-t_2)\right)
%+\arctan\left(\Lambda t_2\right)
%-\frac{\Lambda t_1}{1+\Lambda^{2}t_{2}^{2}}\right)\right\}\notag\\
&&\hspace{-0.4cm}\times\left(ab^{\prime}\rho_{00}(0)e^{i\omega_{S}(t_1-t_2)}+ba^{\prime}\rho_{11}(0)e^{-i\omega_{S}(t_1-t_2)}\right),
\label{two_time CF2}
\end{eqnarray} 
respectively.

\begin{figure}[tbp]
\includegraphics[width=\linewidth]{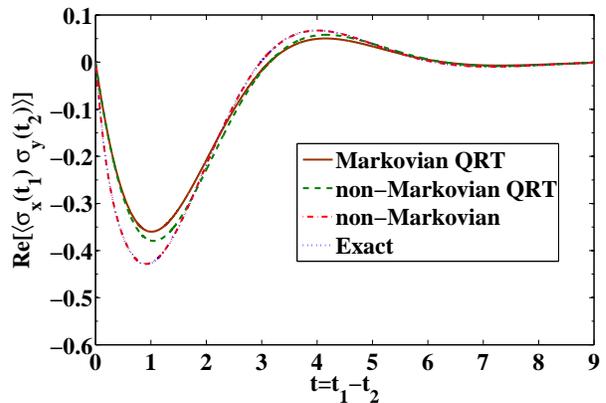}
\caption{(Color online) Time evolutions of the real part of the system operator CF  $\langle\sigma_x(t_1)\sigma_y(t_2)\rangle$  for four different cases:
Markovian (solid line), non-Markovian using the QRT (dashed line) and
non-Markovian (dot-dashed line) using Eq.~(\ref{2time_evol_eq_f})
and exact operator evaluation (dotted line). 
Other parameters used are $\omega %
_{S}=1$, $(k_{B}T/\hbar )=0.1$, $\Lambda =5$, $\gamma =0.1$, and $%
t_{2}=0.2$. }
\label{fig:CF_t2_02} 
\end{figure}

Figure \ref{fig:CF_t2_02} shows the the time evolutions of the real part
of the system operator CF $\langle\sigma_x(t_1)\sigma_y(t_2)\rangle$ 
at a finite temperature of $(k_{B}T/\hbar)=0.1\omega_S$.  
The time evolutions of the CF $\langle\sigma_x(t_1)\sigma_y(t_2)\rangle$ are obtained in four different cases: the first is in the Markovian case, the second is in the non-Markovian case with a finite cut-off frequency but wrongly applying the QRT method [i.e., neglecting the last two terms of 
Eq.~(\ref{2time_evol_eq_f}) or equivalently neglecting the terms with 
$\tilde{D}(t_1,t_2)$ in Eqs.~(\ref{2time_evol_xy})--(\ref{2time_evol_yy})], 
%(which is wrong), 
the third is in the non-Markovian case using the derived evolution equations (\ref{2time_evol_xy})--(\ref{2time_evol_yy}), and the fourth is the exact model using the direct operator evaluation. The initial environment state is in the thermal state and the system state is arbitrarily chosen to be $\vert \Psi\rangle =\left( \frac{\sqrt{3}}{2}\left\vert e\right\rangle +\frac{1}{2}\left\vert g\right\rangle \right)$. 
The Markovian case in Fig. \ref{fig:CF_t2_02} is described as follows.
With the finite cut-off environment spectral density and with the system parameters used in Fig.~\ref{fig:CF_t2_02}, the Markovian approximation may actually 
not be valid. If, however, we still assume that the 
environment correlation time in 
Eq.~(\ref{eq:bathCF}) is much smaller than all of the system time scales 
(i.e., Markovian approximation), 
then we may replace the upper time integration limit in 
Eq.~(\ref{eq:Dt}) to infinity (i.e., $t\to \infty$). 
As a result, the Markovian master equation or evolution equations can be obtained by just replacing the time-dependent coefficient $D(t)$ in Eq.~(\ref{PerturbationME}) or in Eqs.~(\ref{1time_evol_x}) and (\ref{1time_evol_y}) by its long-time limit value. At a finite temperature, the Markovian (time-independent) coefficient from Eqs.~(\ref{eq:Dt}), (\ref{eq:bathCF}) and (\ref{Jw}) can be 
written as  
\begin{equation}
D_\infty=\lim_{t\to\infty}D(t)=4\gamma\pi k_BT/\hbar.  
\label{D_Mar}
\end{equation}
We may see that $D_\infty\to 0$ as the temperature $T\to 0$. This is because in the Markovian limit, the decoherence or dephasing is strongly dependent 
on the infrared behavior ($\omega\to 0$ modes) 
of the environment in the pure dephasing model. 
Since the spectral density  considered in Eq.~(\ref{Jw}) is Ohmic, 
we then have 
$J(\omega\to 0)=0$, and thus $D_\infty\to 0$ at $T=0$. 
This is in contrast to other quantum open system models 
with a resonant type of system-environment coupling, in which the 
environment modes near the system resonance frequency are 
relevant to the relaxation and decoherence.    
We can see from Fig.~\ref{fig:CF_t2_02} that 
the difference between the results of 
the Markovian QRT case and the non-Markovian QRT case is visible, while   
the two-time CF's obtained by the non-Markovian evolution equation (\ref{2time_evol_eq_f}) and by the exact operator evaluation are identical for the pure-dephasing spin-boson model.   
The perfect agreement of the results between the non-Markovian evolution equation case and the exact operator evaluation case, and the
significant difference in the short time region between the non-Markovian evolution equation case and the wrong non-Markovian QRT case 
demonstrate clearly the validity and practical usage of 
the evolution equation (\ref{2time_evol_eq_f}).
%in \cite{Goan09}. 
All of the four cases approach one another to zero in the long time region. 
%where the non-Markovian cases (effects) reduce to the Markovian one. 

\begin{figure}
\includegraphics[width=\linewidth]{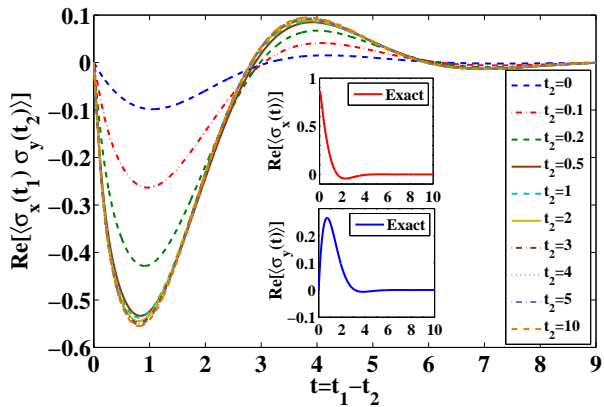}
\caption{(Color online) Time evolutions of the real part of the system operator CF $\langle \sigma_{x}(t_{1})\sigma_{y}(t_{2})\rangle $ for different values of  $t_{2}$
using non-Markovian Eq.~(\ref{2time_evol_eq_f}). The results of the 
time evolutions coincide 
with those obtained by the exact operator evaluation. 
Other parameters used are $\omega _{S}=1$, $%
(k_{B}T/\hbar )=0.1$, $\Lambda =5$, $\gamma =0.1$. 
The insets show the time evolutions of the real part of the 
expectation values 
$\langle \sigma_{x}(t)\rangle $ and 
$\langle \sigma_{y}(t)\rangle $. }
\label{fig:CF_diff_t2}
\end{figure}
\begin{figure}[tbp]
\includegraphics[width=\linewidth]{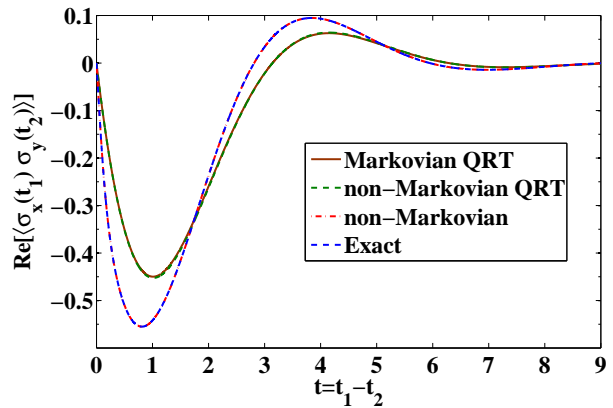}
\caption{(Color online) Time evolutions of the real part of the system operator CF  $\langle\sigma_x(t_1)\sigma_y(t_2)\rangle$  for four different cases:
Markovian (solid line), non-Markovian using the QRT (dashed line) and
non-Markovian (dot-dashed line) using Eq.~(\ref{2time_evol_eq_f})
and exact operator evaluation (dotted line). 
Other parameters used are 
$\omega_{S}=1$, $(k_{B}T/\hbar )=0.1$, $\Lambda =5$, $\gamma =0.1$, and 
$t_{2}=10$. }
\label{fig:CF_t2_10} 
\end{figure}
Figure \ref{fig:CF_diff_t2} investigates the dependence of the exact two-time system operator CF on the time variable $t_ 2$. We see that the time evolutions of the real part of the CF $\langle \sigma_{x}(t_{1})\sigma_{y}(t_{2})\rangle $ as a function of $t=t_1-t_2$ for the values of $t_2\leq 1$ behave quite differently, but they
%the CF's $\langle \sigma_{x}(t_{1})\sigma_{y}(t_{2})\rangle $ 
approach one another for $t_2 \geq 2$. When $t_2\geq 5$, the steady state is reached as indicated in the time evolutions of 
the expectation values $\langle \sigma_{x}(t)\rangle $ and 
$\langle \sigma_{y}(t)\rangle$
shown in the insets of Fig.~\ref{fig:CF_diff_t2}. 
In this case, the time evolutions of the two-time CF 
are independent of the choices of the 
starting time of $t_2$ in the steady state and depend only on the time difference $t=t_1-t_2$ for the parameters used in Fig.~\ref{fig:CF_diff_t2}. 
This can also be seen from the analytical expression of the exact zero-temperature CF (\ref{two_time CF2}). The zero-temperature CF (\ref{two_time CF2}) is a function of variables $t_1$ and $t_2$, but for a large value of the cut-off frequency $\Lambda$, when $t_2$ is reasonably large, the CF depends almost only on $t=t_1-t_2$. 
%One can also see from Fig.~\ref{fig:CF_diff_t2} that the time evolution 
%of the CF when $t_2=0$ is significantly different from that of the 
%steady-state CF (e.g., when $t_2=10$).
Figure \ref{fig:CF_t2_10} shows the time evolutions of the real part of the steady-state ($t_2=10$) system operator CF $\langle\sigma_x(t_1)\sigma_y(t_2)\rangle$ obtained 
in four different cases as in Fig.~\ref{fig:CF_t2_02}. As expected, the non-Markovian evolution equation case coincides with the exact operator evaluation case.
They are, however, significantly different from the wrong non-Markovian QRT 
case and Markovian QRT case, even though the time evolutions of the 
steady-state two-time CF's depend only on the time difference $t=t_1-t_2$. 
One can also see that the CF's of the Markovian QRT and the non-Markovian QRT 
cases approach each other much more closely in the steady state than 
in Fig.~\ref{fig:CF_t2_02}.

\section{Conclusion}
\label{sec:conclusion}
We have evaluated the exact non-Markovian finite-temperature one-time
expectation values
and two-time
CF's of the system operators for the exactly solvable pure-dephasing
spin-boson model. The evaluation has been performed in two ways, one
by exact direct operator technique without any approximation and the
other by the evolution equations 
(\ref{1time_evol_eq_f}) and (\ref{2time_evol_eq_f})
%derived in Ref.~\cite{Goan09} 
valid to second order in the system-environment interaction Hamiltonian. Since the non-Markovian dynamics of the pure-dephasing spin-boson model can be cast into a time-local, convolutionless form and $[L,H_s]=0$,
the results obtained by the second-order evolution equations (\ref{1time_evol_eq_f}) and (\ref{2time_evol_eq_f})
%derived in Ref.~\cite{Goan09} 
turn out to be exactly the same as the exact results obtained by the
exact direct operator evaluation. The agreement of the results between
the two different approaches demonstrates clearly the validity of the
evolution equations 
(\ref{1time_evol_eq_f}) and (\ref{2time_evol_eq_f}).
%derived in Ref.~\cite{Goan09}. 
Furthermore, it is easy 
to obtain Eqs.~(\ref{2time_evol_xz})--(\ref{2time_evol_yz}), Eqs.~(\ref{2time_evol_zx}) and (\ref{2time_evol_zy}), and Eqs.~(\ref{2time_evol_xy})--(\ref{2time_evol_yy}) from the evolution equation (\ref{2time_evol_eq_f}). 
Other non-Markovian open quantum system models that are not exactly solvable can be proceeded in a similar way to obtain the time evolutions of their two-time system operator CF's valid to second order in the 
system-environment interaction Hamiltonian. 
This illustrates the practical usage of the evolution equations.
It is thus believed that the 
evolution equations (\ref{1time_evol_eq_f}) and 
(\ref{2time_evol_eq_f}), 
%or more generally Eqs.~(14) and (17) of Ref.~\cite{Goan09},  
which generalize the QRT 
to the non-Markovian finite-temperature case will have applications in
many different branches of physics.

\begin{acknowledgments}
We would like to acknowledge support from the National Science
Council, Taiwan, under Grant No. 97-2112-M-002-012-MY3, 
support from the Frontier and Innovative Research Program 
of the National Taiwan University under Grants No. 97R0066-65 and 
No. 97R0066-67,
and support from the focus group
program of the National Center for Theoretical Sciences, Taiwan.
We are grateful to the National Center for High-performance Computing, Taiwan, 
for computer time and facilities.
\end{acknowledgments}

\appendix
\section{Derivation of time evolution operator}
To show the time evolution operator of Eq.~(\ref{UI}), 
we begin from Eq.~(\ref{eq:evolution_op}) with $\tilde{H}_I(t)$ given 
by Eq.~(\ref{eq:HIt}).
Since $\tilde{H}_I(t)$ in Eq.~(\ref{eq:HIt}) contains only two major terms, which are, respectively, proportional to $a_{\lambda}$ and $a_\lambda^\dagger$, one is tempting to evaluate the time-ordered exponent 
by the reverse operator identity of Eq.~(\ref{eq:op_id})  
\begin{equation}
e^{A+B}=e^Ae^Be^{-\frac{1}{2}[A,B]},  
\label{eq:op_id_rev}
\end{equation}
valid for the commutator $[A,B]$ commuting with both $A$ and $B$. As the exponent operates at different times and $[\tilde{H}_I(t),\tilde{H}_I(\tau)]\neq 0$, it is not correct to use the precise form of Eq.~(\ref{eq:op_id_rev}). The proper procedure done in \cite{Mahan00,Reina02} is to separate the two terms of Eq.~(\ref{eq:HIt}) in the time-ordered exponent of Eq.~(\ref{eq:evolution_op}) by    
%We have used this formula $:$ $\exp\left(s(A+B)\right)
%=\exp(sA){\rm T}_s\exp\left(\int_{0}^{s}ds^{\prime}
%\exp(-s^{\prime}A)B\exp(s^{\prime}A)\right)$ \cite{Mahan00}. 
\begin{eqnarray}
\tilde{U}(t)&=&\exp\left({-i\int_{0}^{t}d\tau\sum_{\lambda}g_{\lambda}\sigma_{z}e^{i\omega_{\lambda}\tau}a^{\dagger}_{\lambda}}\right)\notag\\
&&\times{\rm T}\left\{\exp\left[-i\int_{0}^{t}d\tau e^{i\int_{0}^{\tau}ds\sum_{\lambda}g_{\lambda}\sigma_{z}e^{i\omega_{\lambda}s}a^{\dagger}_{\lambda}}\right.\right.\notag\\
&&\times\left.\left.\left(\sum_{\lambda}g_{\lambda}\sigma_{z}e^{-i\omega_{\lambda}\tau}a_{\lambda}\right)e^{-i\int_{0}^{\tau}ds\sum_{\lambda}g_{\lambda}\sigma_{z}e^{i\omega_{\lambda}s}a^{\dagger}_{\lambda}}\right]\right\},\notag\\
\label{UITO}
\end{eqnarray} 
where ${\rm T}$ is the time-ordering operator.
Then using the identity 
$e^{-\phi a_\lambda^\dagger}a_\lambda e^{\phi a_\lambda^\dagger}=a_\lambda+\phi$ in the exponent of 
the time-ordered term in Eq.~(\ref{UITO}), we obtain   
\begin{eqnarray}
\tilde{U}(t)&=&\exp\left(-i\int_{0}^{t}d\tau\sum_{\lambda}g_{\lambda}\sigma_{z}e^{i\omega_{\lambda}\tau}a^{\dagger}_{\lambda}\right)\notag\\
&&\times \exp\left(-i\int_{0}^{t}d\tau\sum_{\lambda}g_{\lambda}\sigma_{z}e^{-i\omega_{\lambda}\tau}a_{\lambda}\right)\notag\\
&&\times \exp\left(-\int_{0}^{t}d\tau\int_{0}^{\tau}ds\sum_{\lambda}|g_{\lambda}|^{2}e^{-i\omega_{\lambda}(\tau-s)}\right).
%&=&\exp\left(-i\int\nolimits_{0}^{t}d\tau
%\sum_{\lambda}g_{\lambda}\sigma_z\left(
%e^{i\omega_{\lambda}\tau}a^{\dagger}_{\lambda}
%+e^{-i\omega_{\lambda}\tau}a_{\lambda}\right)\right)\nonumber\\
%&&\times\exp\left(\frac{1}{2}\int_{0}^{t}d\tau\int_{0}^{t}ds\sum_{\lambda}
%|g_{\lambda}|^2 e^{i\omega_{\lambda}(\tau-s)}\right)\nonumber\\
%&&\times\exp\left(-\int_{0}^{t}d\tau\int_{0}^{\tau}ds\sum_{\lambda}
%|g_{\lambda}|^2 e^{-i\omega_{\lambda}(\tau-s)}\right).
\label{UIevaluate}
\end{eqnarray}
We have dropped the time-ordering operator in Eq.~(\ref{UIevaluate}).
Using the operator identity of Eq.~(\ref{eq:op_id}) to combine the first two terms in Eq.~(\ref{UIevaluate}), we then obtain Eq.~(\ref{UI}). Note that the first term in  Eq.~(\ref{UI}) is just $\exp[-(i/\hbar)\int_0^td\tau\tilde{H}_I(\tau)]$ if the time-ordering operator for $\tilde{U}(t)$ is not performed. The correct time-ordering procedure generates extra phase factors in Eq.~(\ref{UIevaluate}) and thus in Eq.~(\ref{UI}).  
%we dropped the time-ordering operator in the second equal sign of 
%Eq.~(\ref{UIevaluate}).

%\section{CONCLUSION}

%\begin{thebibliography}

\end{document}